\documentclass[12pt]{article}

\usepackage{graphicx}
\begin{document}

\begin{center}
{\bf Holographic superconductor with nonlinear Born$-$Infeld-type electrodynamics } \\
\vspace{5mm} S. I. Kruglov
\footnote{E-mail: serguei.krouglov@utoronto.ca}
\underline{}
\vspace{3mm}

\textit{ Department of Physics, University of Toronto, \\60 St. Georges St.,
Toronto, ON M5S 1A7, Canada\\
Department of Chemical and Physical Sciences,\\ University of Toronto Mississauga,\\
3359 Mississauga Rd. N., Mississauga, ON L5L 1C6, Canada}
\vspace{5mm}
\end{center}

\begin{abstract}
Holographic s-wave superconductors in the framework of nonlinear Born$-$Infeld-type electrodynamics is investigated
in the background of Schwarzschild anti-de Sitter black holes. As particular cases, at some model parameters, we obtain results for Born$-$Infeld and exponential electrodynamics. We explore the analytical Sturm-Liouville eigenvalue problem in the probe limit where the scalar and electromagnetic fields do not effect on the background metric. The critical temperatures of phase transitions and the order parameter are calculated which depend on the model parameters. We show that the critical exponent near the critical temperature is 1/2.  Making use of the matching method we derive analytical expressions for the condensation values and the critical temperature. The conductivity by the analytical method is calculated.
\end{abstract}

\section{Introduction}

The Maldacena conjecture \cite{Maldacena} gives the connection of type-IIB string theory with a conformal field theory (N = 4 SU(N) supersymmetric Yang-Mills theory). This correspondence between anti-de Sitter (AdS) spacetime and the conformal field
theory (CFT) on a boundary was obtained from the string theory \cite{Maldacena}, \cite{Gubser}, \cite{Witten}.
This duality (AdS/CFT correspondence) was proven also for other gravitational backgrounds and without supersymmetry and conformal symmetry.
The AdS/CFT correspondence is a strong/weak duality connecting the strongly coupled quantum field theory and the weakly interacting gravitational theory. In other words this holographic duality allows us to investigate strongly interacting systems by studying dynamical fields in the bulk gravitational theory. Thus, the infrared physics of the bulk gravitational theory near the boundary has a counterpart in the ultraviolet physics of dual CFT. Instead of studying a strongly interacting field theory of condensed matter system in $d$ dimensions one can investigate a classical gravity in $d+1$ dimensions \cite{Hartnoll}, \cite{Hartnoll_1}, \cite{Herzog}.
This duality may help to understand the mechanism of high temperature superconductors in condensed matter physics \cite{Hartnoll2}.
When gravity is coupled to gauge fields then charged scalars can condense near the horizon
and black hole horizons cause to spontaneous breaking of an Abelian gauge symmetry  \cite{Gubser_1}, \cite{Gubser_2}.
The thermodynamics of CFT corresponds to the thermodynamics of the black hole in the dual gravitational theory, and the Hawking temperature of the black hole is the same as the temperature in the CFT. The entropy of the black hole, connected with the black hole horizon, is equal to
the thermal entropy of CFT. Some s-wave holographic superconductors were investigated in \cite{Umeh}-\cite{Cheng}. The holographic superconductors in the framework of Born-Infeld (BI) nonlinear electrodynamics (NLED) were studied in \cite{Roychowdhury_0}, \cite{Sunandan}, \cite{Lai}, \cite{Jing_1}, \cite{Roychowdhury}, \cite{Ghoraia1}, \cite{Cheng}.

Here, we study holographic s-wave superconductors with B-I-type electrodynamics proposed in \cite{Kruglov}, \cite{Kruglov_0}.
In this model of NLED the electric field of a point-like object is finite at the center and the static electric self-energy is finite. The same properties hold in BI electrodynamics. Within BI-type electrodynamics the regular black hole solution was obtained in \cite{Kruglov_0}. The four-dimensional AdS$_4$ can provide an analog to the phenomena in the thin-film superconductors.
We explore the Sturm-Liouville analytical approach and we are restricted by the probe limit.

The paper is organised as follows. In section 2, we consider the field equations of AdS$_4$ black holes with BI-type electrodynamics. The critical temperature is evaluated as a function of the charge density in section 3. In section 4, we find
the condensates of the scalar operators.  Analytical expressions for the condensation values and the critical temperature are found with the help of the matching method in section 5. We evaluate the conductivity with the aid of the analytical method in section 6. The results are discussed in section 7.

\section{Field equations for scalar and gauge fields}

Let us investigate the formation of a scalar hair of the AdS$_4$ black hole for a holographic s-wave superconductor in the probe limit.
We use the planar line element of the background, for the Schwarzschild-AdS$_4$ black hole, in the form \cite{Hartnoll}
\begin{equation}
ds^2=-f(r)dt^2+\frac{1}{f(r)}dr^2+r^2(dx^2+dy^2),
\label{1}
\end{equation}
where the metric function of the black hole is defined by
\begin{equation}
f(r)=\frac{r^2}{L^2}-\frac{M}{r}.
\label{2}
\end{equation}
Here $L$ is the curvature radius of the AdS$_4$ black hole and the negative cosmological constant equals $\Lambda=-3/L^2$. The asymptotically AdS$_4$ region takes place at large $r$. Roots of the equation $f(r_+)=0$ determine a regular
black hole horizon radius $r_+$ so that $r_+=(ML^2)^{1/3}$. One can obtain the Hawking temperature from Eq. (2)
\begin{equation}
T_H=\frac{\kappa}{2\pi}=\frac{f'(r_+)}{4\pi}=\frac{3M^{1/3}}{4\pi L^{4/3}},
\label{3}
\end{equation}
where $\kappa$ is the surface gravity. We explore the Lagrangian density of
BI-type electrodynamics (${\cal L}_0$), proposed in \cite{Kruglov}, \cite{Kruglov_0}, and the Lagrangian density of a charged complex scalar field
\begin{equation}
{\cal L} ={\cal L}_0 -|(\nabla_\mu-iqA_\mu)\psi|^2-V(|\psi|),
 \label{4}
\end{equation}
\begin{equation}
{\cal L}_0 = \frac{1}{\beta}\left[1-\left(1+\frac{\beta{\cal F}}{\sigma}-\frac{\beta\gamma {\cal G}^2}{2\sigma}\right)^\sigma\right],
 \label{5}
\end{equation}
where the background is defined by the metric function (2). Here ${\cal F}=-(1/4)F^{\mu\nu}F_{\mu\nu}=(\textbf{B}^2-\textbf{E}^2)/2$, $F_{\mu\nu}=\nabla_\mu A_\nu-\nabla_\nu A_\mu$ is the field strength tensor, and ${\cal G}=(1/4)F_{\mu\nu}\tilde{F}^{\mu\nu}=\textbf{E}\cdot \textbf{B}$, $\tilde{F}^{\mu\nu}=(1/2)\epsilon^{\mu\nu\alpha\beta}F_{\alpha\beta}$. The parameter $\beta$ has the dimensions of (length)$^4$, and $\sigma$ ($-\infty<\sigma<\infty$) is the dimensionless parameter. If $0<\sigma<1$ the maximum of the electric field at the center of charges is $E(0)=\sqrt{2\sigma/\beta}$ \cite{Kruglov}, \cite{Kruglov_0}.
At $\sigma=1$, $\gamma=0$ we come from Eq. (5) to classical electrodynamics and
when $\sigma=1/2$ the model (1) becomes the model proposed in \cite{Krug}.
If $\beta=\gamma$ and $\sigma=1/2$ the model (5) is converted into BI electrodynamics \cite {Born} and at the limit $\sigma\rightarrow \infty$ NLED (5) becomes exponential electrodynamics, ${\cal L}_{exp}= (1/\beta)\left[1-\exp\left(\beta{\cal F}-\beta\gamma {\cal G}^2/2\right)\right]$.

It should be noted that the correspondence principle takes place, i.e. at the weak field limit, $\beta {\cal F}\ll 1$, $\gamma{\cal G}\ll 1$, the model (5) is converted into Maxwell's electrodynamics, ${\cal L}_0\rightarrow -{\cal F}$. In the following we put $\gamma=0$.
The potential function $V(|\psi|)$ of the scalar field $\psi$ is used in the form  \cite{Hartnoll}
\begin{equation}
V(|\psi|) =m^2|\psi|^2,
 \label{6}
\end{equation}
where $m^2=-2/L^2$ and it is above the Breitenlohner-Freedman bound \cite{Breitenlohner}. We imply the probe limit i.e. the field $\psi$
is small and does not back-react on the geometry. At $\beta\rightarrow 0$ the Lagrangian density (5) becomes the Maxwell Lagrangian density ${\cal L}_M=-{\cal F}$. We are going to study the condensate formation in the dual theory (CFT) and to obtain solutions of scalar fields that do not back-react. From Eqs. (4), (5) and (6) one finds field equations
\begin{equation}
\frac{1}{\sqrt{-g}}\partial_\mu\left(\frac{\sqrt{-g}F^{\mu\nu}}{(1+\beta{\cal F}/\sigma)^{1-\sigma}}\right)=iq\left(\psi^*\partial^\nu\psi
-\psi\partial^\nu\psi^*\right)+2q^2A^\nu|\psi|^2,
 \label{7}
\end{equation}
\[
\partial_\mu\left(\sqrt{-g}\partial^\mu\psi\right)-iq\sqrt{-g}A^\mu\partial_\mu\psi-iq\partial_\mu(\sqrt{-g}A^\mu\psi)-q^2\sqrt{-g}A_\mu A^\mu \psi
\]
\begin{equation}
-\sqrt{-g}m^2\psi=0.
 \label{8}
\end{equation}
Let us assume that the scalar field $\psi$ and the Abelian gauge field $A_\mu$ depend only on $r$ \cite{Hartnoll} as follows:
\begin{equation}
A_\mu=(\phi(r),0,0,0)),~~~~\psi=\psi(r).
 \label{9}
\end{equation}
We use a choice of units $q=1$ which corresponds to rescaling the fields $\psi$ and $A_\mu$ \cite{Hartnoll}. The phase of scalar fields can be put to zero, i.e. the field $\psi$ is real. With the help of Eqs. (7), (8) and (9) one obtains
\begin{equation}
\frac{1}{r^2}\partial_r\left(\frac{r^2\phi'(r)}{\left(1-\beta\phi^{'2}(r)/(2\sigma)\right)^{1-\sigma}}\right)-
\frac{2\psi^2(r)}{f(r)}\phi(r)=0,
 \label{10}
\end{equation}
\begin{equation}
\psi''(r)+\left(\frac{2}{r}+\frac{f'(r)}{f(r)}\right)\psi'(r)+\frac{\phi^2(r)}{f^2(r)}\psi(r)+\frac{2}{L^2f(r)}\psi(r)=0.
 \label{11}
\end{equation}
If $\beta\rightarrow 0$ we arrive at Maxwell's electrodynamics and Eqs. (10) and (11) become equations studied in \cite{Hartnoll}.
We also put $L=1$ and $r_+=M^{1/3}$ and then the parameter $\beta$ becomes unitless. At the horizon, $r=r_+$, $\phi=0$, and Eq. (11) gives to the equality $\psi=-3r_+\psi'/2$.
Solutions to Eqs. (10) and (11) at $r\rightarrow\infty$ behave as
\begin{equation}
\phi(r)=\mu-\frac{\rho}{r}+ {\cal O}(r^{-2}),
 \label{12}
\end{equation}
\begin{equation}
\psi=\frac{\psi^{(1)}}{r}+\frac{\psi^{(2)}}{r^2}+{\cal O}(r^{-3}),
 \label{13}
\end{equation}
where $\mu$ is the chemical potential and $\rho$ is the charge density of the dual field theory.
The values $\psi^{(i)}$ ($i=1,2$) are interpreted as condensates of the scalar operators ${\cal O}_i$ in the dual theory, $\psi^{(i)}=\langle{\cal O}_i\rangle/\sqrt{2}$ \cite{Hartnoll}.
The asymptotic of AdS$_4$ is stable if $\psi^{(1)}\neq 0$, $\psi^{(2)}=0$ or $\psi^{(1)}=0$, $\psi^{(2)}\neq0$.
We use the choice $\psi^{(2)}=0$ and $\psi^{(1)}\neq 0$.

\section{Critical temperature of holographic superconductor}

To study the dependence of the critical temperature on the charge density we integrate
 Eq. (10) at $\psi=0$, and we find the equation
\begin{equation}
C\left(1-\beta\frac{\phi^{'2}(r)}{2\sigma}\right )^{1-\sigma}=r^2\phi'(r),
 \label{14}
\end{equation}
where $C$ is a constant of integration. Equation (14) shows that at $0<\sigma<1$ the electric field $E=-\phi'(r)$ at the center of an object possesses the maximum finite value $E(0)=\sqrt{2\sigma/\beta}$ \cite{Kruglov}. Similar behavior holds in BI nonlinear electrodynamics \cite{Born}. The asymptotic series of the function $\phi'(r)$ in Eq. (14) at $r\rightarrow \infty$ (or in the small parameter $\beta$) is
\begin{equation}
\phi'(r)=\frac{C}{r^2}-\frac{(1-\sigma)C^3\beta}{2\sigma r^{6}} +{\cal O}(\beta^2r^{-8}).
 \label{15}
\end{equation}
We assume that $(1-\sigma)C^3\beta/(2\sigma r^6)\ll 1$.
After integration of Eq. (15) and using a new variable $z=r_+/r$, one obtains
\begin{equation}
\phi(z)=\int_{r_{+(c)}}^r \phi'(r)dr=\frac{\rho}{r_{+(c)}}(1-z)+\frac{(1-\sigma)\rho^3\beta(z^5-1)}{10\sigma r_{+(c)}^{5}} +{\cal O}(\beta^{2}),
 \label{16}
\end{equation}
where $C\equiv\rho$. We use in Eq. (16) $r_+=r_{+(c)}$ as solution (16) was obtained at $\psi=0$ and corresponds to $T=T_c$.
As a result, the corrections to classical electrodynamics are in the order of ${\cal O}(\beta^2)$.  At $z = 1$ we have the horizon $r=r_+$ and
the value $z = 0$ corresponds to the boundary $r\rightarrow \infty$. The metric function is given by $f(z)=r_+^2(1-z^3)/z^2$, and Eq. (11) becomes
\begin{equation}
\psi''(z)-\frac{2+z^3}{z(1-z^3)}\psi'(z)+\left[\frac{\phi^2(z)}{r_+^2(1-z^3)^2}+\frac{2}{z^2(1-z^3)}\right]\psi(z)=0.
 \label{17}
\end{equation}
From Eq. (16) we obtain
\begin{equation}
\phi^2(z)= \lambda^2r_{+(c)}^2(z-1)^2\left(1-\frac{(1-\sigma)\lambda^2\beta\xi(z)}{5\sigma}\right)+{\cal O}(\beta^{2}),
 \label{18}
\end{equation}
where the notations $\lambda=\rho/r_{+(c)}^2$ and $\xi(z)=(z^5-1)/(z-1)$ are used. Near the critical point $T\approx T_c$, and from equations (17) and (18) up to ${\cal O}(\beta^{2})$, one finds
\[
\psi''(z)-\frac{2+z^3}{z(1-z^3)}\psi'(z)+\frac{2}{z^2(1-z^3)}\psi(z)
\]
\begin{equation}
+\frac{\lambda^2}{(1+z+z^2)^2}\left[1-\frac{(1-\sigma)\lambda^2\beta\xi(z)}{5\sigma}\right]\psi(z)=0.
 \label{19}
\end{equation}
We define the scalar field  $\psi(z)$ at the boundary in the form
\begin{equation}
\psi(z)=\frac{\langle{\cal O}_1\rangle}{\sqrt{2}r_+}zF(z).
 \label{20}
\end{equation}
The trial function is chosen as $F(z)=1-\alpha z^2$ with the boundary conditions $F(0)=1$, $F'(0)=0$ \cite{Siopsis}.
Replacing Eq. (20) into Eq. (19) one obtains the differential equation
\begin{equation}
\left[(1-z^3)F'(z)\right]'-zF(z)+\frac{\lambda^2(1-z)}{1+z+z^2}\left[1-\frac{(1-\sigma)\lambda^2\beta\xi(z)}{5\sigma}\right]F(z)=0.
 \label{21}
\end{equation}
 Equation (21)can be represented in the Sturm-Liouville form
\begin{equation}
\left[p(z)F'(z)\right]'+q(z)F(z)+\lambda^2g(z)F(z)=0,
 \label{22}
\end{equation}
with functions
\begin{equation}
p(z)=1-z^3,~~q(z)=-z,~~g(z)=\frac{(1-z)}{1+z+z^2}\left[1-\frac{(1-\sigma)\lambda^2\beta\xi(z)}{5\sigma}\right].
 \label{23}
\end{equation}
The minimum of the eigenvalue squared, $\lambda^2$, can be evaluated from the expression \cite{Gangopadhyay}
\begin{equation}
\lambda^2=\frac{\int_0^1\{p(z)[F'(z)]^2-q(z)[F(z)]^2\}dz}{\int_0^1g(z)[F(z)]^2dz}.
 \label{24}
\end{equation}
In Eq. (23), for the function $g(z)$, one can use terms which are linear in the parameter $\beta$ because $\beta\lambda^2=\beta\lambda^2|_{\beta=0}+{\cal O}(\beta^2)$.
 Making use of Eqs. (23) and (24) we obtain the expression
\[
\lambda^2=\frac{5\alpha^2-3\alpha+3}{(6\alpha^2+6\alpha-3)\ln(3)-6.5\alpha^2-18\alpha+\sqrt{3}\pi(2\alpha+1)+\varepsilon \lambda^2|_{\beta=0} B},
\]
\begin{equation}
B=-\frac{337}{700}\alpha^2+2.88\alpha+1.1-0.6\ln(3)(\alpha^2+4\alpha+1)+0.2\sqrt{3}\pi(\alpha^2-1),
 \label{25}
\end{equation}
where $\varepsilon=(1-\sigma)\beta/\sigma$. Then the critical temperature (3) at $L=1$ and $M=r_+^3$ becomes
\begin{equation}
T_c=\frac{3}{4\pi}\sqrt{\frac{\rho}{\lambda}}.
 \label{26}
\end{equation}
By minimizing the value $\lambda^2$ in Eq. (24) with respect to $\alpha$, one finds, according to Eq. (26), the maximum value of the critical temperature $T_c$, and can obtain $T_c$ from Eq. (26).

In Table 1 we represent the critical temperatures for different parameters $\varepsilon$. For BI electrodynamics ($\sigma=0.5$, $\varepsilon=\beta$) our results for $\beta\equiv b=0, 0.1, 0.2, 0.3$ are in agreement with the results obtained in \cite{Roychowdhury}.
\begin{table}[ht]
\caption{Critical temperature}
\centering
\begin{tabular}{c c c c c c c c c c}\\[1ex]
 \hline
$\varepsilon$ & 0 & 0.1 & 0.2 & 0.3 & 0.4 & -0.1 & -0.2 & -0.3 & -0.4\\[1ex]
\hline 
 $\alpha$ & 0.2389 & 0.2402 & 0.2414 & 0.2427 & 0.2440 & 0.2376 & 0.2363 & 0.2351 & 0.2338 \\[1ex]
\hline
 $\lambda^2_{min}$ & 1.2683 & 1.3172 & 1.3699 & 1.4271 & 1.4892 & 1.2230 & 1.1807 & 1.1413 & 1.1044 \\[1ex]
\hline
 $T_c/\sqrt{\rho}$ & 0.2250 & 0.2228 & 0.2207 & 0.2184 & 0.2161 & 0.2270 & 0.2290 & 0.2310 & 0.2329\\[1ex]
\hline
\end{tabular}
\end{table}
It follows from Table 1 that if the parameter $\beta$ increases the critical temperature decreases. The critical temperature for BI electrodynamics ($\varepsilon=\beta>0$) is smaller compared to BI-type electrodynamics at $\varepsilon<0$ ($\sigma>1$ ). Then to create the condensation with BI-type electrodynamics at $\sigma>1$ is easier as compared with BI and classical electrodynamics.

\section{Condensates and the critical exponent}

Condensates in the dual theory can be evaluated by considering Eq. (10) near the critical point. Equation (10) with variable $z=r_+/r$ can be represented as follows:
\[
\left(1-\frac{(2\sigma-1)\beta z^4}{2\sigma r_+^2}\phi^{'2}(z)\right)\phi''(z)+\frac{2(1-\sigma)\beta z^3}{\sigma r_+^2}\phi^{'3}(z)
\]
\begin{equation}
=\frac{2\phi(z)\psi^2(z)}{z^2(1-z^3)}\left(1-\frac{\beta z^4}{2\sigma r_+^2}\phi^{'2}(z)\right)^{2-\sigma}.
 \label{27}
\end{equation}
By placing Eq. (20) into Eq. (27) and using the expansion of the right side of Eq. (27) in the small parameter $\beta$, one obtains
\[
\left(1-\frac{(2\sigma-1)\beta z^4}{2\sigma r_+^2}\phi^{'2}(z)\right)\phi''(z)+\frac{2(1-\sigma)\beta z^3}{\sigma r_+^2}\phi^{'3}(z)
\]
\begin{equation}
=\frac{\langle{\cal O}_1\rangle F^2(z)\phi(z)}{r_+^2(1-z^3)}\left(1-\frac{(2-\sigma)\beta z^4}{2\sigma r_+^2}\phi^{'2}(z)\right).
 \label{28}
\end{equation}
As $\langle{\cal O}_1\rangle^2/r_+^2\ll 1$, we can expand $\phi(z)$ as follows:
\begin{equation}
\frac{\phi(z)}{r_+}=\frac{\phi_0(z)}{r_+}+\frac{\langle{\cal O}_1\rangle^2}{r_+^2}\chi(z)+\cdot\cdot\cdot,
 \label{29}
\end{equation}
where $\phi_0(z)$ obeys the equation
\begin{equation}
\left(1-\frac{(2\sigma-1)\beta z^4}{2\sigma r_+^2}\phi^{'2}_0(z)\right)\phi''_0(z)+\frac{2(1-\sigma)\beta z^3}{\sigma r_+^2}\phi^{'3}_0(z)
=0.
 \label{30}
\end{equation}
Replacing Eq. (29) into Eq. (28), and neglecting small parameters, one finds the equation
\[
\chi''(z)\left(1-\frac{(2\sigma-1)\beta z^4}{2\sigma r_+^2}\phi_0^{'2}(z)\right)+\frac{6(1-\sigma))\beta z^3\phi^{'2}_0(z)}{\sigma r_+^2}\chi'(z)
\]
\begin{equation}
=\frac{ F^2(z)\phi_0(z)}{r_+^2(1-z^3)}\left(1-\frac{(2-\sigma)\beta z^4}{2\sigma r_+^2}\phi^{'2}(z)\right).
 \label{31}
\end{equation}
The function $\chi(z)$ satisfies the boundary conditions $\chi(1)=\chi'(1)=0$. Then the approximate solution to Eq. (30) is
\begin{equation}
\phi_0(z)=\frac{\rho(1-z)}{r_{+(c)}} +\frac{(1-\sigma)\rho^3\beta(z^5-1)}{10\sigma r_{+(c)}^5}+{\cal O}(\beta^2).
 \label{32}
\end{equation}
By replacing Eq. (32) into Eq. (31), one obtains (up to $\beta^2$) the equation
\[
\left(1-\frac{(2\sigma-1)\beta\lambda^2z^4}{2\sigma}\right)\chi''(z)+\frac{6(1-\sigma)\beta\lambda^2z^3}{\sigma}\chi'(z)
\]
\begin{equation}
=\frac{F^2(z)\lambda}{1+z+z^2}\left(1-\frac{2-\sigma}{2\sigma}\beta\lambda^2z^4-\frac{1-\sigma}{10\sigma}\beta\lambda^2\xi\right).
 \label{33}
\end{equation}
Here we use the notations $\lambda=\rho/r_{+(c)}^2$ and $\xi(z)=(z^5-1)/(z-1)$. Making use of the expansion of Eq. (33) in small parameters, it becomes
\begin{equation}
\zeta'(z)+P(z)\zeta(z)=Q(z),
 \label{34}
\end{equation}
where
\[
\zeta(z)=\chi'(z),~~P(z)=\frac{6(1-\sigma)\beta\lambda^2z^3}{\sigma},
\]
\begin{equation}
Q(z)=\frac{(1-\alpha z^2)^2\lambda}{1+z+z^2}\left(1+\frac{3(\sigma-1)}{2\sigma}\beta\lambda^2z^4-\frac{1-\sigma}{10\sigma}\beta\lambda^2\xi\right),
 \label{35}
\end{equation}
and $F(z)=1-\alpha z^2$.
 The solution to Eq. (34) is given by
\begin{equation}
\zeta(z)=\exp(-I(z))\left(\int Q(z)\exp(I(z))dz+C\right),
 \label{36}
\end{equation}
with $C$ being the constant of integration, and
\begin{equation}
I(z)=\int P(z)dz=\frac{3(1-\sigma)}{2\sigma}\beta\lambda^2z^4.
 \label{37}
\end{equation}
By expanding the exponential function in small parameter $\beta$, one finds
\begin{equation}
\int Q(z)\exp(I(z))dz=\lambda\int\frac{(1-\alpha z^2)^2}{1+z+z^2}\left(1-\frac{1-\sigma}{10\sigma}\beta\lambda^2\xi(z)\right)dz+{\cal O}(\beta^2),
\label{38}
\end{equation}
Let us use the boundary condition
$\chi'(1)=0$ that gives $\zeta(1)=0$, and with the help of Eqs. (36), (38) and the condition $\zeta(1)=0$, neglecting ${\cal O}(\beta^2)$, one obtains the integration constant
\[
C=-\lambda\biggl\{\frac{1}{6}\biggl[-\alpha(\alpha+12)-2\sqrt{3}(\alpha^2-2\alpha-2)\arctan\sqrt{3}+3\alpha(\alpha+2)\ln(3)\biggr]
\]
\begin{equation}
-\frac{1-\sigma}{10\sigma}\lambda^2\beta\left[\frac{25}{28}\alpha^2+\frac{1}{\sqrt{3}}(\alpha^2-2\alpha-2)\arctan\sqrt{3}
-\frac{7}{5}\alpha+\frac{1}{3}+\frac{2\alpha+1}{2}\ln(3)\right]\biggr\}.
 \label{39}
\end{equation}
Making use of Eqs. (36), (38) and (39) we find
\begin{equation}
\chi'(0)=\zeta(0)=-\lambda {\cal A},
 \label{40}
 \end{equation}
 \[
{\cal A}=\frac{\alpha^2-2\alpha-2}{\sqrt{3}}\biggl(\arctan\frac{1}{\sqrt{3}}-(\arctan\sqrt{3}\biggr)-\frac{1}{6}\alpha(\alpha+ 12)+\frac{1}{2}\alpha(\alpha+2)\ln(3)
\]
\[
+\frac{1-\sigma}{10\sigma}\lambda^2\beta\biggl[(2\alpha^2+2\alpha-1)\left(\arctan\sqrt{3}-\arctan\frac{1}{\sqrt{3}}\right)
\]
\begin{equation}
-\frac{2\alpha+1}{2}\ln(3)-\frac{25}{28}\alpha^2
+\frac{7}{5}\alpha-\frac{1}{3}\biggr].
 \label{41}
 \end{equation}
By using the procedure \cite{Roychowdhury}, with the help of Eqs. (12) and (29), taking into account Eqs. (26), (32) and (40),
 $T=3r_+/(4\pi)$ and  series $\chi(z)=\chi(0)+z\chi'(0)+\cdot\cdot\cdot$, we obtain the order parameter
\begin{equation}
\langle{\cal O}_1\rangle=\gamma T_c\sqrt{1-\frac{T}{T_c}},
 \label{42}
 \end{equation}
 \begin{equation}
\gamma=\frac{4\pi\sqrt{2}}{3\sqrt{{\cal A}}}.
 \label{43}
 \end{equation}
The values $\gamma$ for different parameters $\varepsilon$ are given in Table 2.
\begin{table}[ht]
\caption{Condensation values $\gamma$}
\centering
\begin{tabular}{c c c c c c c c c c}\\[1ex]
 \hline
$\varepsilon$ & 0 & 0.1 & 0.2 & 0.3 & 0.4 & -0.1 & -0.2 & -0.3 & -0.4 \\[1ex]
\hline 
 $\alpha$ & 0.2389 & 0.2402 & 0.2414 & 0.2427 & 0.2440 & 0.2376 & 0.2363 & 0.2351 & 0.2338 \\[1ex]
\hline
 $\gamma$ & 8.074 & 8.185 & 8.309 & 8.450 & 8.611 & 7.975 & 7.885 & 7.803 & 7.729 \\[1ex]
\hline
\end{tabular}
\end{table}
Our values for $\gamma$ in the case of BI electrodynamics ($\sigma=0.5$, $\varepsilon=\beta$) for $\beta\equiv b=0, 0.1, 0.2, 0.3$ are in agreement with \cite{Roychowdhury}.
The plot of order parameter $\langle{\cal O}_1\rangle/T_c$ versus $T/T_c$ is depicted in Fig. 1.
\begin{figure}[h]
\includegraphics[height=3.0in,width=3.0in]{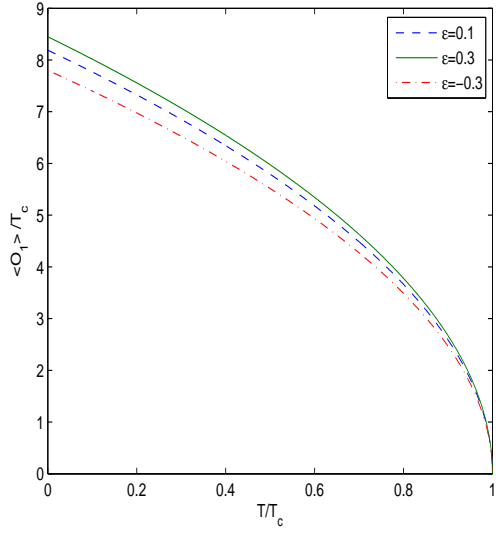}
\caption{\label{fig.1}The plot of the order parameter $\langle{\cal O}_1\rangle/T_c$ vs. $T/T_c$.}
\end{figure}

\section{Condensation values with matching method}

We will derive analytic expressions for the condensation value and the critical temperature with the help of the matching method \cite{Soda}. Making use of Taylor series, near the horizon, one finds
\begin{equation}
\phi(z)=\phi(1)+\phi'(1)(z-1)+\frac{1}{2}\phi''(1)(z-1)^2,
 \label{44}
 \end{equation}
 \begin{equation}
\psi(z)=\psi(1)+\psi'(1)(z-1)+\frac{1}{2}\psi''(1)(z-1)^2.
 \label{45}
 \end{equation}
We obtain the equation from Eq. (27) up to ${\cal O}(\beta^2)$
\[
\phi''(z)+\frac{2(1-\sigma)\beta z^3\phi^{'3}(z)}{\sigma r_+^2}
-\frac{2\phi(z)\psi^2(z)}{z^2(1-z^3)}\left(1+\frac{3(\sigma-1)\beta z^4\phi^{'2}(z)}{2\sigma r_+^2}\right)
\]
\begin{equation}
+{\cal O}(\beta^2)=0.
 \label{46}
\end{equation}
From Eqs. (44), (46) and the boundary condition $\phi(1)=0$, near $z = 1$, one finds
\begin{equation}
\phi''(1)=-\frac{2(1-\sigma)\beta\phi^{'3}(1)}{\sigma r_+^2}
-\frac{2\phi'(1)\psi^2(1)}{3}\left(1+\frac{3(\sigma-1)\beta\phi^{'2}(1)}{2\sigma r_+^2}\right)+{\cal O}(\beta^2).
 \label{47}
 \end{equation}
Replacing Eq. (47) in Eq. (44) we obtain
\[
\phi(z)=\phi'(1)(z-1)-\frac{1}{2}(z-1)^2\biggl[\frac{2(1-\sigma)\beta\phi^{'3}(1)}{\sigma r_+^2}
\]
\begin{equation}
+\frac{2\phi'(1)\psi^2(1)}{3}\left(1+\frac{3(\sigma-1)\beta\phi^{'2}(1)}{2\sigma r_+^2}\right)\biggr]+{\cal O}(\beta^2).
 \label{48}
 \end{equation}
From Eqs. (17), (44) and using the boundary condition $\psi'(1)=2\psi(1)/3$, near $z = 1$, one finds
\begin{equation}
\psi''(1)=-\frac{\phi^{'2}(1)\psi(1)}{9r_+^2}.
 \label{49}
 \end{equation}
Replacing (49) into (45) and taking into account $\psi'(1)=2\psi(1)/3$ we obtain
\begin{equation}
\psi(z)=\frac{1}{3}\psi(1)(1+2z)-\frac{\phi^{'2}(1)\psi(1)}{18r_+^2}(z-1)^2.
 \label{50}
 \end{equation}
Making use of the matching method of \cite{Soda} (see also \cite{Roychowdhury_0}), we can obtain an analytical expression for the
critical temperature $T_c$. By matching asymptotic solutions (12), (13) with (48) and (50) at  $z = z_m$ one finds
\[
\mu-\frac{\rho z_m}{r_+}=\phi'(1)(z_m-1)-\frac{1}{2}(z_m-1)^2\phi'(1)\biggl[\frac{2(1-\sigma)\beta\phi^{'2}(1)}{\sigma r_+^2}
\]
\begin{equation}
+\frac{2\psi^2(1)}{3}\left(1+\frac{3(\sigma-1)\beta\phi^{'2}(1)}{2\sigma r_+^2}\right)\biggr],
 \label{51}
 \end{equation}
\begin{equation}
\frac{\psi^{(1)}z_m}{r_+}=\frac{1}{3}\psi(1)\left(1+2z_m\right)-\frac{\phi^{'2}(1)\psi(1)}{18r_+^2}(z_m-1)^2.
 \label{52}
 \end{equation}
We need the equality of derivatives at $z = z_m$ to match these asymptotic solutions smoothly
\[
\frac{\rho}{r_+}=-\phi'(1)+(z_m-1)\phi'(1)\biggl[\frac{2(1-\sigma)\beta\phi^{'2}(1)}{\sigma r_+^2}
\]
\begin{equation}
+\frac{2\psi^2(1)}{3}\left(1+\frac{3(\sigma-1)\beta\phi^{'2}(1)}{2\sigma r_+^2}\right)\biggr],
 \label{53}
 \end{equation}
\begin{equation}
\frac{\psi^{(1)}}{r_+}=\frac{2}{3}\psi(1)-\frac{\phi^{'2}(1)\psi(1)}{9r_+^2}(z_m-1).
 \label{54}
 \end{equation}
Making use of notations $b = - \phi'(1)$, $a = \psi(1)$ and Eq. (53) one finds
\begin{equation}
a^2=\frac{3}{2(1-z_m)}\left[\frac{\rho}{br_+}-1-\beta\frac{(1-\sigma)b^2}{\sigma r_+^2}\left(\frac{7}{2}-\frac{3\rho}{2br_+}-2z_m\right)\right]+{\cal O}(\beta^2).
 \label{55}
 \end{equation}
Taking into account of $T\equiv T_H=3r_+/(4\pi)$, from (55) we obtain
\begin{equation}
a^2=\frac{3}{2(1-z_m)}\left(\frac{T_c}{T}\right)^2\left(1+\frac{(1-\sigma)\beta\tilde{b}^2(7-4z_m)}{2\sigma}\right)\left(1-\frac{T^2}{T_c^2}\right)+{\cal O}(\beta^2),
 \label{56}
 \end{equation}
where $\tilde{b}=b/r_+$. The critical temperature is given by
\begin{equation}
T_c=\frac{3\sqrt{\rho}}{4\pi\sqrt{\tilde{b}}}\sqrt{1-\frac{2(1-\sigma)\beta \tilde{b}^2(1-z_m)}{\sigma}}.
 \label{57}
 \end{equation}
Near the critical temperature ($T \approx T_c$), from (56) one finds
\begin{equation}
a=\sqrt{\frac{3}{1-z_m}}\left(1+\frac{(1-\sigma)\beta\tilde{b}^2(7-4z_m)}{4\sigma}\right)\sqrt{1-\frac{T}{T_c}}+{\cal O}(\beta^2).
 \label{58}
 \end{equation}
From Eqs. (52) and (54) we obtain
\begin{equation}
\tilde{b}=\sqrt{\frac{6}{1-z_m^2}},~~~~\psi^{(1)}=\frac{2ar_+(2+z_m)}{3(1+z_m)}.
 \label{59}
 \end{equation}
The order parameter $\langle{\cal O}_1\rangle=\sqrt{2}\psi^{(1)}$, near the critical temperature $T \approx T_c$ found from Eqs. (58) and (59)  becomes
\[
\langle{\cal O}_1\rangle=\frac{8\sqrt{2}\pi}{3\sqrt{3}}\left(\frac{2+z_m}{1+z_m}\right)\sqrt{\frac{1}{1-z_m}}
\left(1+\frac{3(1-\sigma)\beta(7-4z_m)}{2\sigma(1-z_m^2)}\right)
\]
\begin{equation}
\times T_c\sqrt{1-\frac{T}{T_c}}+{\cal O}(\beta^2).
 \label{60}
 \end{equation}
We obtain from Eq. (57) the upper bound on the parameter $\beta$,
\begin{equation}
\beta^\leq \frac{\sigma(1+z_m)}{12(1-\sigma)}.
 \label{61}
 \end{equation}
The critical temperature becomes lower due to nonlinear corrections, and it is harder to have the
scalar condensate at low temperature. In Fig. 2 the critical temperature $T_c$ vs. $\rho$ is depicted for
$z_m = 1/2$.
\begin{figure}[h]
\includegraphics[height=3.0in,width=3.0in]{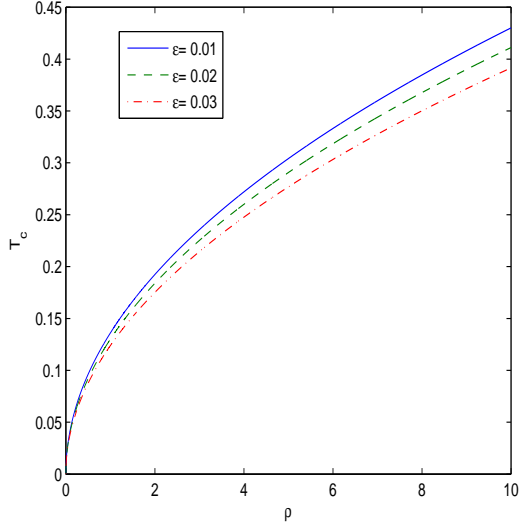}
\caption{\label{fig.2}The plot of critical temperature $T_c$ vs. $\rho$ for different choices of the parameter $\varepsilon=(1-\sigma)\beta/\sigma$ ($z_m=0.5$).}
\end{figure}
The plot of $T/\sqrt{\rho}$ vs. the matching parameter $z_m$ is given in Fig. 3.
\begin{figure}[h]
\includegraphics[height=3.0in,width=3.0in]{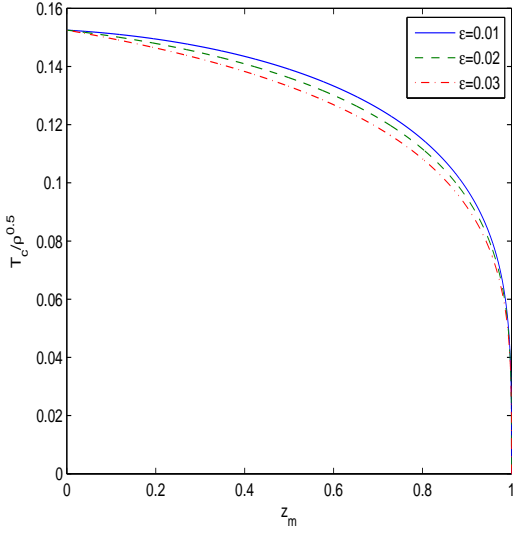}
\caption{\label{fig.3}The $T_c/\sqrt{\rho}$ vs. the matching parameter $z_m$.}
\end{figure}
The plot of the condensation values $\langle{\cal O}_1\rangle/T_c$ vs. $T/T_c$ ($z_m=0.5$) is represented in Fig. 4.
\begin{figure}[h]
\includegraphics[height=3.0in,width=3.0in]{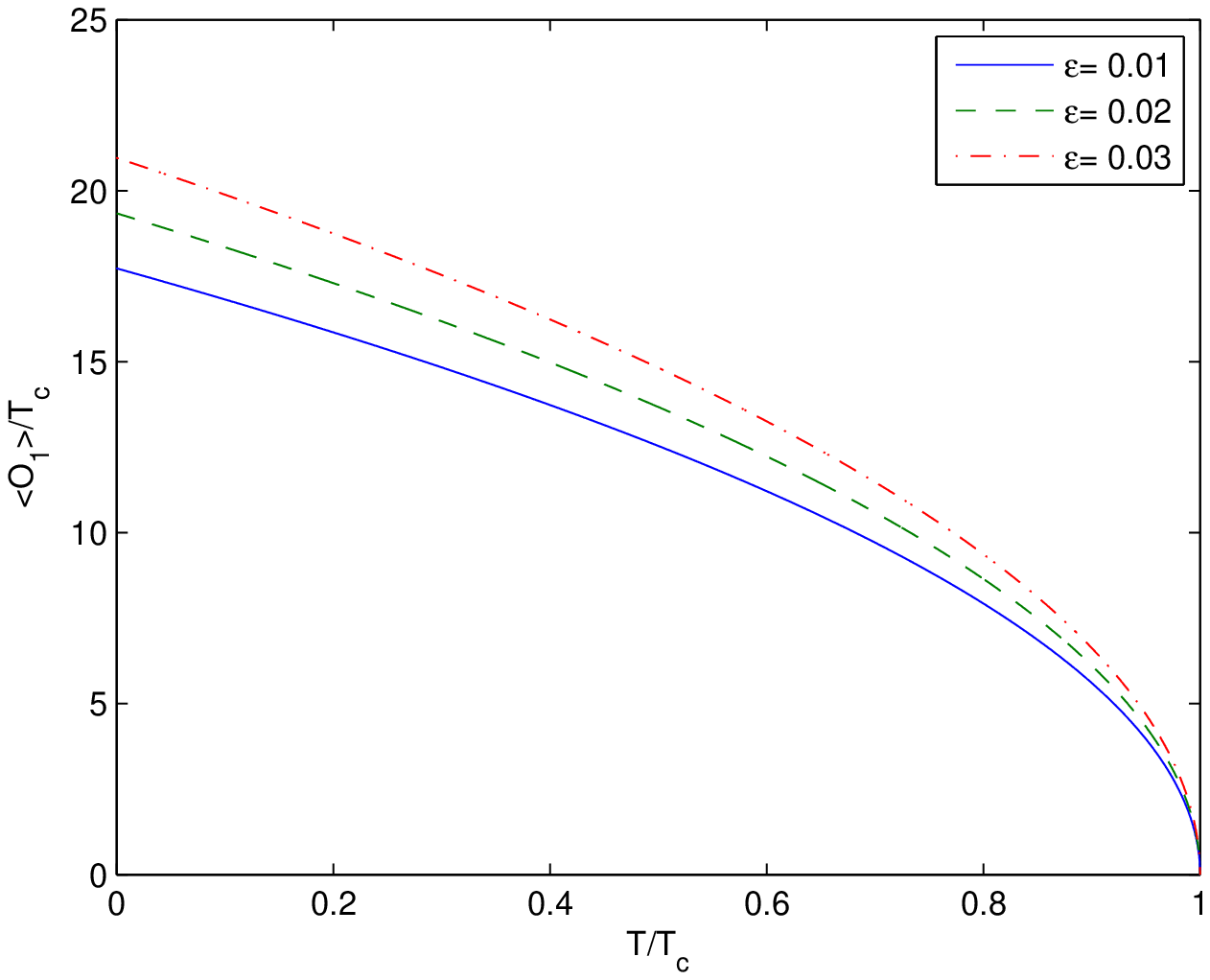}
\caption{\label{fig.4}The condensation values $\langle{\cal O}_1\rangle/T_c$ vs. $T/T_c$ ($z_m=0.5$).}
\end{figure}

According to Fig. 3 the different choices of $z_m$, close to $z_m\approx 0.5$, does not give a big difference in numerical values for the critical temperature and the condensation values.

\section{Conductivity}

Let us calculate the low temperature conductivity in the dual theory. One has to obtain the  vector-potential $A_x(t,r)$ making use of an electromagnetic perturbation. Then we find the conductivity on the $AdS$ boundary. With the help of Eq. (7) we obtain the equation along the boundary
\begin{equation}
\frac{1}{r^2}\partial_0\left(\frac{r^2F^{0x}}{(1+\beta{\cal F}/\sigma)^{1-\sigma}}\right)+\frac{1}{r^2}\partial_r\left(\frac{r^2F^{rx}}{(1+\beta{\cal F}/\sigma)^{1-\sigma}}\right)=2A^x\psi^2,
 \label{62}
\end{equation}
with ${\cal F}=-E_xE^x/2=-(\partial_0A_x)\partial_0A^x$. Using sinusoidal electromagnetic perturbations, $A_x\sim\exp (-i\omega t)$, and equalities $A^x=A_x/r^2$, $F^{0x}=-\partial_0 A_x/(f(r)r^2)$, $F^{rx}=f(r)(\partial_r A_x)/r^2$, from Eq. (62) one finds
\begin{equation}
\partial_0\left(\frac{i\omega A_x}{f(r)(1+\beta\omega^2 A^2/(2\sigma))^{1-\sigma}}\right)
+\partial_r\left(\frac{f(r)\partial_rA_x}{(1+\beta\omega^2 A^2/(2\sigma))^{1-\sigma}}\right)=2A_x\psi^2,
 \label{63}
\end{equation}
where $A^2=A_xA^x$. Making use of Eq. (63) we obtain
\[
\frac{\omega^2A_x}{f(r)}+\frac{(\sigma-1)\omega^4\beta A_xA^2}{\sigma f(r)[1+\beta\omega^2 A^2/(2\sigma)]}+\left(f(r)A'_x\right)'
+\frac{\omega^2\beta f(r)A'_x(A^2)'}{2\sigma[1+\beta\omega^2 A^2/(2\sigma)]}
\]
\begin{equation}
=2A_x\psi^2[1+\beta\omega^2 A^2/(2\sigma)]^{1-\sigma},
 \label{64}
\end{equation}
where $A'_x =\partial_r A_x$. Using Taylor series in small parameter $\beta$ one fonds
\[
\left(f(r)A'_x)\right)'+\left(\frac{\omega^2}{f(r)}-2\psi^2\right)A_x
\]
\begin{equation}
=\varepsilon\omega^2\left(2A_xA^2\psi^2+\frac{2\omega^2 A_xA^2}{f(r)}+f(r)A'_x\left(A^2\right)'\right)+{\cal O}(\varepsilon^2),
 \label{65}
\end{equation}
where $\varepsilon=(1-\sigma)\beta(2\sigma)$.
At $\beta=0$ we come to the equation obtained in \cite{Hartnoll}. To find a solution to Eq. (65) we use the equation
\begin{equation}
A_x= \bar{A_x}+\varepsilon B,
 \label{66}
\end{equation}
where $\bar{A_x}$ obeys the homogenous equation
 \begin{equation}
\left(f(r)\bar{A_x}')\right)'+\left(\frac{\omega^2}{f(r)}-2\psi^2\right)\bar{A_x}=0.
 \label{67}
\end{equation}
Making use of Eqs. (65) and (66) one finds the equation
\[
\left(f(r)B')\right)'+\left(\frac{\omega^2}{f(r)}-2\psi^2\right)B
\]
\begin{equation}
=\omega^2\left(2\bar{A_x}\bar{A}^2\psi^2+\frac{2\omega^2 \bar{A_x} \bar{A}^2}{f(r)}+f(r)\bar{A_x}'\left(\bar{A}^2\right)'\right)+{\cal O}(\varepsilon^2).
 \label{68}
\end{equation}
At $r\rightarrow \infty$ we have $f(r)\approx r^2$, $\psi\approx \langle {\cal O}_1\rangle/(\sqrt{2}r)$, near the boundary, and the solution to Eq. (67) is \cite{Hartnoll}
 \begin{equation}
\bar{A_x}=\exp\left( i\frac{\sigma_0\omega}{r}\right),
 \label{69}
\end{equation}
with the conductivity for $\beta=0$
\begin{equation}
\sigma_0=\sqrt{1-\frac{\langle {\cal O}_1\rangle^2}{\omega^2}}.
 \label{70}
\end{equation}
Equation (70) is valid for low temperatures and the limit $M\rightarrow 0$.
Replacing Eq. (69) into Eq. (68), and using Eq. (70), we obtain
\[
\left(r^2B')\right)'+\frac{B}{2r^2}\left(2\omega^2-\langle {\cal O}_1\rangle^2\right)
\]
\begin{equation}
=\frac{\omega^2}{r^4}\left[\frac{\langle {\cal O}_1\rangle^2}{2}+2\omega^2+2i\sigma_0\omega\left(r+ i\sigma_0\omega\right)\right]\exp\left(\frac{3i\sigma_0\omega}{r}\right)+{\cal O}(\varepsilon^2).
 \label{71}
\end{equation}
One can find the solution to Eq. (71) as follows:
\begin{equation}
B=\left(c_0+\frac{c_1}{r}+\frac{c_2}{r^2}\right)\exp\left(\frac{3i\sigma_0\omega}{r}\right)+{\cal O}(\varepsilon^2),
 \label{72}
\end{equation}
with
\[
c_0=\frac{4(156\sigma_0^3-258\sigma_0^2-85\sigma_0^4-5)}{(1-17\sigma_0^2)^3},
\]
\begin{equation}
c_1=\frac{4i\omega\sigma_0(13\sigma_0-29)}{(1-17\sigma_0^2)^2},~~~~
c_2=\frac{5\omega^2(1-\sigma_0^2)}{1-17\sigma_0^2}.
 \label{73}
\end{equation}
By using Eqs. (66), (69) and (72) at large radius, we obtain
\begin{equation}
A_x=1+\frac{i\sigma_0\omega}{r}+\varepsilon\left(c_0+\frac{c_1}{r}+\frac{3i\sigma_0\omega c_0}{r}\right)+{\cal O}(r^{-2}).
 \label{74}
\end{equation}
Making use of equations \cite{Hartnoll}
\begin{equation}
\sigma(\omega)=-\frac{iA_x^{(1)}}{\omega A_x^{(0)}},~~~A_x=A_x^{(0)}+\frac{A_x^{(1)}}{r}+{\cal O}(r^{-2}),
\label{75}
\end{equation}
where according to the AdS/CFT correspondence, $A_x^{(0)}$ is the source and $A_x^{(1)}$ is dual to the
current, and taking into account of Eqs. (73),(74) and (75), one finds the conductivity
\[
\sigma(\omega)=\sigma_0+\varepsilon\left(2\sigma_0c_0-\frac{ic_1}{\omega}\right)+{\cal O}(\varepsilon^2)
\]
\begin{equation}
=\sigma_0+\varepsilon\frac{4\sigma_0}{\left(1-17\sigma_0^2\right)^3}\left(91\sigma_0^3-23\sigma_0^2-170\sigma_0^4+13\sigma_0-39\right)+{\cal O}(\varepsilon^2).
\label{76}
\end{equation}
The conductivity $\sigma_0$ is in good agreement with numerical results in high and low frequencies at $\omega\ll \langle {\cal O}_1\rangle$
 \cite{Hartnoll}. Corrections to $\sigma_0$, due to BI-type electrodynamics are in the order of ${\cal O}(\varepsilon^2)$. The conductivity (76) is real at high frequencies $\omega>\langle {\cal O}_1\rangle$ ($\sigma=\mbox{Re}[\sigma]$). At low frequencies $\omega<\langle {\cal O}_1\rangle$ one can find the imaginary and real parts of the conductivity. It should be noted that the conductivity (76) has  poles at $\omega=0$ and at $1-17\sigma_0^2=0$ ($\omega=\sqrt{17/16}\langle{\cal O}_1\rangle\approx 1.03\langle{\cal O}_1\rangle$). The corrections to the conductivity $\sigma_0$, due to the presence of nonlinear terms in BI-type electrodynamics, have to be small. Therefore, the pole $\omega=\sqrt{17/16}\langle{\cal O}_1\rangle$ is non-physical and the usage of $\sigma$ close to this pole is questionable.
The plot of $\mbox{Re}[\sigma]$ at $\omega>\langle {\cal O}_1\rangle$ versus $\omega/\langle {\cal O}_1\rangle$ is depicted in Fig. 5.
\begin{figure}[h]
\includegraphics[height=3.0in,width=3.0in]{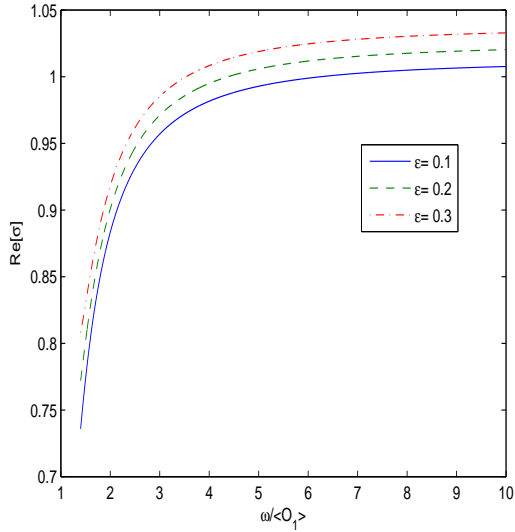}
\caption{\label{fig.5} The real part of the conductivity vs. $\omega/\langle{\cal O}_1\rangle$.}
\end{figure}
Fig. 5 shows that with the increasing $\varepsilon$ the conductivity Re$[\sigma_0]$ also increasing.

The imaginary part of the conductivity $\sigma$ at $\omega=0$ has a pole, Im$[\sigma(\omega)]=\langle{\cal O}_1\rangle/\omega$ and corrections to the conductivity, due to nonlinear terms, do not contribute to this pole. The real part of the conductivity, with the help of the Kramers-Kronig relations, has a delta function, Re$[\sigma(\omega)]=\pi \langle{\cal O}_1\rangle\delta(\omega)$ \cite{Hartnoll} leading to a DC superconductivity. This DC superconductivity is due to a second order phase transition.
The real part of the conductivity versus $\omega/T$ at $\varepsilon=0.2$ can be found from Eqs. (42) and (76) and is represented in Fiq. 6.
\begin{figure}[h]
\includegraphics[height=3.0in,width=3.0in]{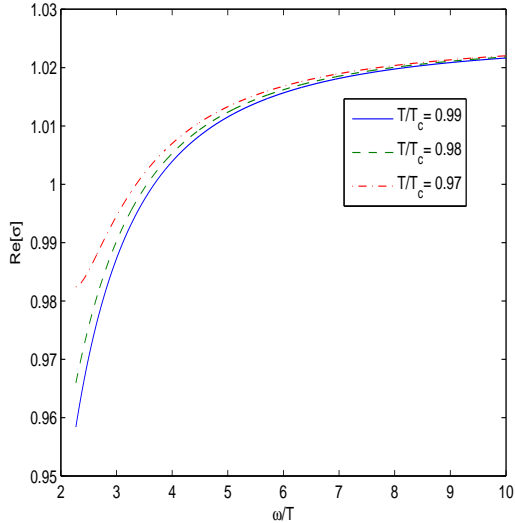}
\caption{\label{fig.6} The real part of the conductivity vs. $\omega/T$ at $\varepsilon=0.2$.}
\end{figure}
The formation of a gap in the real part of the conductivity is shown in Fig. 6. The normal phase takes place at high frequencies and at low frequencies, below the critical temperature, the condensate is formed.

 \section{Conclusion}

 The holographic s-wave superconductors with BI-type electrodynamics in the presence of Schwarzschild AdS$_4$ background was studied. We explored the analytical approach for the Sturm-Liouville eigenvalue problem bacause the numerical calculations are not accurate when the temperature approaches to zero \cite{Hartnoll2}, \cite{Siopsis}.
 The probe limit was used so that the scalar and electromagnetic fields do not effect on the background in the bulk theory. The condensation operator and critical temperature, near the critical point, depend on the parameters $\beta$ and $\sigma$ of the model. By increasing the parameter $\varepsilon=(1-\sigma)\beta/\sigma$ the critical temperature decreases. The same behaviour takes place in the holographic s-wave superconductors with BI electrodynamics. It is interesting that the condensates become smaller and the critical temperature is greater (see Tables 1 and 2) with $\sigma>1$ ($\varepsilon<0$) as compared with BI electrodynamics and classical electrodynamics. In our model the critical exponent of holographic s-wave superconductors is equal to $1/2$ as well as in other models. The order parameter depending on $\beta$ and $\sigma$ was calculated. Making use of the matching method we obtained analytical expressions
for the condensation values and the critical temperature. It should be noted that the critical temperature and the condensation values depend on the $z_m$ weakly. The method based on the Sturm-Liouville eigenvalue problem and the matching method give similar results.

\end{document}